\documentclass[sigconf]{acmart}
\AtBeginDocument{%
  }
\setcopyright{acmlicensed}

\usepackage{amsmath}
\usepackage{amsmath} 
\usepackage{subcaption}
\usepackage{listings}
\usepackage{booktabs}
\usepackage{multirow}
\usepackage{array}
\usepackage{geometry}
\usepackage{siunitx}
\usepackage{makecell}
\usepackage{adjustbox}
\usepackage[table]{xcolor}
\definecolor{best}{RGB}{225,245,225}
\definecolor{worst}{RGB}{255,230,230}

\newcolumntype{P}[1]{>{\centering\arraybackslash}p{#1}}

\usepackage{tikz}

\usepackage{tcolorbox}
\newtcolorbox{boxK}{
    fontupper = \small,
    sharpish corners, 
    boxrule = 0pt,
    toprule = 0pt, 
}

\newcommand{\realsearchgoal}{\textit{The goal of this study is to aid security researchers in extracting attack techniques and controls from a campaign by replicating and comparing the performance of the state-of-the-art ATT\&CK technique extraction methods in a multi-report campaign setting compared to prior single-report evaluations. }}

\usepackage{amsthm}

\newtheoremstyle{uprightstyle}%
  {3pt}{3pt}{\normalfont}{}{\bfseries}{.}{ }{}
\theoremstyle{uprightstyle}

\newtheorem{researchquestion}{RQ}

\newcommand{\storeRQ}[2]{%
  \expandafter\newcommand\csname #1\endcsname{#2}%
}

\newcommand{\printRQ}[2]{%
  \begin{researchquestion}\label{#2}%
    \csname #1\endcsname
  \end{researchquestion}%
}

\newcommand{\reuseRQ}[1]{%
  \textbf{\autoref{#1}:} 
}

\storeRQ{RQEffectiveness}{How do existing automated attack technique extraction methods perform at the campaign level, having multiple reports compared to a single report?}

\storeRQ{RQErrorAnalysis}{Which attack techniques are most frequently missed or misclassified by existing methods, and does semantic similarity between ATT\&CK technique descriptions contribute to these errors?}

\storeRQ{RQMitigationGap}{What is the impact of automated attack technique extraction performance on the coverage of mitigation controls?}

\storeRQ{RQSaturation}{How many CTI reports are required for automated approaches to reach performance saturation as additional reports are incorporated?} 
\storeRQ{RQCTICharacteristics}{What characteristics of CTI reports enable existing attack technique extraction approaches to perform better?}

\usepackage{circledsteps}
\usepackage{tikz}
\usepackage{amsmath}
\usepackage{booktabs}
\usepackage{multirow}
\usepackage{array}
\usepackage{siunitx}
\usepackage{caption}
\usepackage{hyperref}

\usepackage{cleveref}
\usepackage{multirow}
\usepackage{booktabs}
\usepackage{xcolor}
\usepackage{colortbl}
\usepackage{arydshln}

\usetikzlibrary{shapes.geometric, arrows.meta, positioning, fit, backgrounds, calc}

\definecolor{phasecolor}{HTML}{1F4E79} 
\definecolor{rq1color}{HTML}{2E7D32}   
\definecolor{rq2color}{HTML}{B9770E}   
\definecolor{rq3color}{HTML}{5E35B1}   
\definecolor{rq4color}{HTML}{00838F}   
\definecolor{rq5color}{HTML}{C2185B}   

\newcommand{\best}[1]{#1}
\definecolor{bestcolor}{RGB}{255, 255, 204}  
\newcommand{\bestcell}[1]{\cellcolor{bestcolor}\best{#1}}

\copyrightyear{2026}
\acmYear{2026}
\setcopyright{cc}
\setcctype{by}
\acmConference[ASE '26]{2026 IEEE/ACM International Conference on Software Engineering}{October 12--16, 2026 }{Munich, Germany}

\renewcommand\footnotetextcopyrightpermission[1]{}

\begin{document}

\title{Beyond Single Reports: Evaluating Automated ATT\&CK Technique Extraction in Multi-Report Campaign Settings}

\author{Md Nazmul Haque}
\email{mhaque4@ncsu.edu}
\affiliation{%
  \institution{North Carolina State University}
  \city{Raleigh}
  \state{North Carolina}
  \country{USA}
}

\author{Sivana Hamer}
\email{sahamer@ncsu.edu}
\affiliation{%
  \institution{North Carolina State University}
  \city{Raleigh}
  \state{North Carolina}
  \country{USA}
}

\author{Brandon Wroblewski}
\email{bnwroble@ncsu.edu}
\affiliation{%
  \institution{North Carolina State University}
  \city{Raleigh}
  \state{North Carolina}
  \country{USA}
}

\author{Md Rayhanur Rahman}
\email{mdrayhanur.rahman@ua.edu}
\affiliation{%
  \institution{University of Alabama}
  \city{Tuscaloosa}
  \state{Alabama}
  \country{USA}
}

\author{Laurie Williams}
\email{lawilli3@ncsu.edu}
\affiliation{%
  \institution{North Carolina State University}
  \city{Raleigh}
  \state{North Carolina}
  \country{USA}
}
\renewcommand{\shortauthors}{Haque et al.}




\begin{abstract}
Large-scale cyberattacks, referred to as campaigns, are documented across multiple CTI reports from diverse sources, with some providing a high-level overview of attack techniques and others providing technical details. 
Extracting attack techniques from reports is essential for organizations to identify the controls required to protect against attacks. Manually extracting techniques at scale is impractical. Existing automated methods focus on single reports, leaving many attack techniques and their controls undetected, resulting in a fragmented view of campaign behavior.
\realsearchgoal 
We conduct an empirical study of 29 methods to extract attack techniques, spanning entity recognition (NER), encoder-based classification, and decoder-based LLM approaches. Our study analyzes 90 CTI reports across three major attack campaigns, SolarWinds, XZ Utils, and Log4j, using both quantitative performance metrics and their impact on controls. Our results show that aggregating multiple CTI reports improves the F1 score by $\approx26\%$ over single-report analysis, with most approaches reaching performance saturation after 5–15 reports. Despite these gains, extraction performance remains limited, with maximum F1 scores of 78.6\% for SolarWinds and 54.9\% for XZ Utils. Moreover, up to 33.3\% of misclassifications involve semantically similar techniques that share tactics and overlap in descriptions. The misclassification has a disproportionate effect on control coverage. Reports that are longer and include technical details consistently perform better, even though their readability scores are low.
Based on the findings, we advocate that researchers move beyond single-report evaluations and instead use our performance saturation and control coverage metrics to evaluate technique-extraction methods in multi-report campaigns.
\end{abstract}
\maketitle

\vspace{-2mm}
\section{Introduction}
\label{sec:introduction}
Cyber Threat Intelligence (CTI) reports are a key source for understanding cyber attacks, capturing attack techniques (e.g., credential dumping), identifying exploited vulnerabilities, and recommending controls to mitigate the attack~\cite{bromiley2016threat, johnson2016guide}. With the global cost of cyberattacks projected to reach \$12.2 trillion annually by 2031~\cite{2025cybercrime}, organizations increasingly rely on CTI to anticipate, detect, and respond to cyber threats~\cite{chen2025aecr}.
Large-scale cyber attacks are often structured as attack campaigns, defined as coordinated sequences of attack techniques targeting specific organizations or sectors over time~\cite{mitre_attack_campaign}. As a result, multiple CTI reports are often published for the same campaign by different sources~\cite{hamer2026closing}. For example, the SolarWinds campaign is documented by organizations, ranging from government agencies such as CISA to independent security researchers and incident response teams.
While these sources describe the same campaign, they do so from different perspectives: one report may focus on high-level strategic goals, while another provides a granular forensic analysis of a specific malware payload~\cite{hamer2026closing}.

At the same time, according to an IBM survey, organizations receive an average of 60,000 security blog posts per month~\cite{ibmreport}. As the volume and complexity of CTI reports grow, the timely extraction and structuring of attack techniques becomes critical for anticipating and responding to threats~\cite{al2024mitre}. To structure the information, the MITRE ATT\&CK framework~\cite{mitre_attack} has emerged as the de facto standard. For example, \textit{phishing} is an attack technique mapped to ATT\&CK ID T1566. Organizations extract attack techniques from CTI reports and map them to ATT\&CK techniques and use the resulting intelligence to anticipate, detect, and respond to attacks~\cite{mitre_attack, tounsi2019cyber}. However, manually mapping CTI reports to MITRE ATT\&CK framework is a time-consuming and error-prone task~\cite{chen2025aecr, buchel2025sok}, motivating automated extraction approaches.

Prior works have proposed different automated approaches on single-report datasets, including rule-based Named Entity Recognition (NER)~\cite{gao2021enabling, husari2017ttpdrill, li2022attackg, liao2016acing, satvat2021extractor}, encoder-based classification~\cite{huang2024mitretrieval, engenuity2023threat, orbinato2022automatic, rahman2024alert, rani2023ttphunter}, and decoder-based LLM approaches~\cite{chen2025aecr, cheng2024ctinexus, fayyazi2024advancing, fieblinger2024actionable, siracusano2023time}. However, all these approaches use custom datasets and settings, which makes them incomparable. Büchel et al.~\cite{buchel2025sok} conducted a systematic comparison with a unified dataset and evaluation framework to directly compare the performance of different approaches. Even the best-performing approach on single-report datasets achieves a maximum F1-score of 72.5\%, leaving a substantial portion of ATT\&CK techniques undetected. And when an ATT\&CK technique is missed, the corresponding controls, defined as recommended actions that protect against specific ATT\&CK techniques, may also be missed, leaving the organization exposed to undetected threats. While prior work refers to controls as ``mitigations'' or ``tasks,'' we use the term ``controls'' consistently. As such, the performance of the current extraction method, considering a single report, is insufficient for practical use. Moreover, a recent study has demonstrated that aggregating reports of related incidents improves automated methods for understanding software vulnerabilities~\cite{anandayuvaraj2024fail}, suggesting that a similar aggregation strategy may benefit multi-report analyses of the same campaign. Prior work also notes that individual CTI reports may cover different aspects of the same campaign, with one report complementing another in describing attack techniques and campaign behavior~\cite{hamer2026closing}. Aggregating multiple reports for a campaign would help automated methods more accurately identify the ATT\&CK techniques. Hence, accurately capturing ATT\&CK techniques also improves coverage of the controls that protect against them.

\realsearchgoal
In this study, we address the following research questions.

\printRQ{RQEffectiveness}{RQEffectiveness}

\printRQ{RQErrorAnalysis}{RQErrorAnalysis}

\printRQ{RQMitigationGap}{RQMitigationGap}

\printRQ{RQSaturation}{RQSaturation}

\printRQ{RQCTICharacteristics}{RQCTICharacteristics}

To answer the research questions, we conduct a conceptual replication~\cite{dennis2015replication} and extension~\cite{carver2010towards} of Büchel et al.~\cite{buchel2025sok}. We evaluate the same 29 state-of-the-art automated extraction methods spanning three approaches (NER, encoder-based classification, and decoder-based LLM) in a campaign-level, multi-report setting. Rather than evaluating each CTI report in isolation, as prior benchmarks do, we aggregate predictions across multiple reports describing the same attack campaign to assess extraction performance at the campaign level. We conduct this evaluation on a dataset of 90 CTI reports drawn from three high-profile campaigns: SolarWinds, XZ Utils, and Log4j~\cite{hamer2026closing}. 


We found that aggregating multiple CTI reports improves ATT\&CK technique coverage compared to single-report evaluations by $\approx26\%$. 
Our error analysis reveals that 33.3\% of misclassifications occur between semantically similar ATT\&CK techniques, with approximately 79.2\% of these errors involving techniques that share the same tactic (e.g., Defense Evasion, Discovery). We further observe that extraction errors propagate to downstream controls to mitigate the attack techniques, where the best-performing method correctly covers only 77.1\% of the ground-truth controls.
Finally, we found that most approaches reach performance saturation after incorporating 10–15 CTI reports, with technically dense reports from sources such as MITRE and CISA consistently contributing the most to extraction performance.



The main contributions of this study are as follows: 
\Circled{1} A campaign-level replication study of existing automated extraction methods across multiple CTI reports of the same campaign.
\Circled{2} Metrics for evaluating the performance of automated CTI extraction approaches at the control level, quantifying both technique coverage and missed controls.
\Circled{3} Empirical saturation thresholds quantifying the minimum number of CTI reports required for automated extraction approaches to reach performance saturation across attack campaigns.
\Circled{4} A characterization of the textual and structural report properties that maximize ATT\&CK technique extraction performance, providing actionable guidance for cost-effective CTI collection and curation.
\Circled{5} A systematic characterization of the ATT\&CK techniques most frequently missed or misclassified by existing methods to determine whether semantic overlap drives these errors.

The rest of the paper is organized as follows: \S~\ref{sec:related_work} provides the review of the background and related works, \S~\ref{sec:methodology} describes the overall methodology of our study, \S~\ref{sec:results} and \S~\ref{sec:discussion} discuss the results and discussions and \S~\ref{sec:threats_to_validity} and \S~\ref{sec:conclusion} discuss the threats to validity and conclude our paper, respectively.

\vspace{-2mm}
\section{Background and Related Work}
\label{sec:related_work}
Existing automated extraction of ATT\&CK techniques from CTI reports can be broadly categorized into three extraction approaches~\cite{huang2024mitretrieval, buchel2025sok}: \Circled{1} Named Entity Recognition (NER), \Circled{2} encoder-based classification, and \Circled{3} decoder-based LLM. 

\Circled{1} \textbf{Named-Entity Recognition (NER):} NER-based approach treats ATT\&CK technique extraction as a token-level sequence labeling task,  identifying explicit attack techniques within CTI reports and then applying rule-based or machine learning techniques to map extracted phrases to ATT\&CK techniques. A typical NER pipeline includes five components for extracting relevant tokens: tokenization~\cite {webster1992tokenization}, POS tagging~\cite{manning2011part}, lemmatization~\cite{plisson2004rule, balakrishnan2014stemming}, related-word detection~\cite{fellbaum1998wordnet}, and parsing~\cite{parsing}. For example, Husari et al. proposed TTPDril, which used tokenization, POS tagging, and BM25 TF-IDF method to extract relevant information from the CTI reports~\cite{husari2017ttpdrill}. Similarly, Rahman et al. applied BM25 TF-IDF with subject-verb-object tuple extraction~\cite{rahman2022threat}. AttacKG~\cite{li2022attackg} constructs knowledge graphs using NER by extracting attack-relevant entities.

\Circled{2} \textbf{Encoder-based classification:} Encoder-based classification approach formulates attack technique extraction as a multi-class or multi-label text classification task. Instead of identifying specific keywords or tokens, individual sentences are mapped to one or more ATT\&CK techniques. Most of the existing literature uses either BERT- or RoBERTa-based pretrained encoder models to extract ATT\&CK techniques. For instance, researchers~\cite{mitre_tram_2023, rani2023ttphunter, rani2024ttpxhunter} use different variants of BERT-based models~\cite{ranade2021cybert, bayer2024cysecbert, jin2023darkbert, aghaei2022securebert, beltagy2019scibert}. MITREtrieval ~\cite{huang2024mitretrieval} uses a RoBERTa-based model~\cite{liu2019roberta}. While prior work evaluates both variants on individual reports, our study evaluates them at the campaign level, aggregating predictions across multiple CTI reports for the same attack campaign.

\Circled{3} \textbf{Decoder-based LLM:} Decoder-based LLM approach uses LLM to generate attack technique predictions directly from CTI text, typically through instruction-following or prompt-based generation. Instead of relying on predefined labels or token-level supervision, decoder-based LLMs synthesize explanations, infer latent adversarial behaviors, and map narrative descriptions to corresponding ATT\&CK techniques. Instead of classifying raw text into predefined ATT\&CK techniques from a given prompt, the decoder-based LLM approach generates raw text, which is then post-processed using regular expressions to extract ATT\&CK techniques~\cite{siracusano2023time}. These approaches often leverage few-shot prompting (FSP)~\cite{cheng2024ctinexus, fieblinger2024actionable, fengrui2024few, kumarasinghe2024semantic} or retrieval-augmented generation (RAG)~\cite{chen2025aecr, xu2024intelex, cheng2024ctinexus, fayyazi2024advancing} to improve extraction from large or complex CTI reports. Several studies also employ supervised fine-tuning (SFT)~\cite{chen2025aecr, fengrui2024few, fieblinger2024actionable} to adapt LLMs to domain-specific CTI data. For example, AECR fine-tuned a 6-billion-parameter model to outperform larger general-purpose models and reduce hallucinations using a linear classification head~\cite{chen2025aecr}. In contrast, other works, such as CTINexus~\cite{cheng2024ctinexus} and IntelEX~\cite{xu2024intelex}, combine generative LLMs with multi-agent or RAG-like architectures to improve entity relationship extraction and precision. AttacKG+~\cite{zhang2025attackg+} uses a four-stage generative LLM pipeline to extract TTPs and entity relationships to build a knowledge graph.
Anandayuvaraj et al. propose an LLM-based pipeline that collects and groups news articles describing the same software failure incident~\cite{anandayuvaraj2024fail}. While both works aggregate information across multiple documents, their focus is on software incident analysis rather than campaign-level ATT\&CK technique extraction.

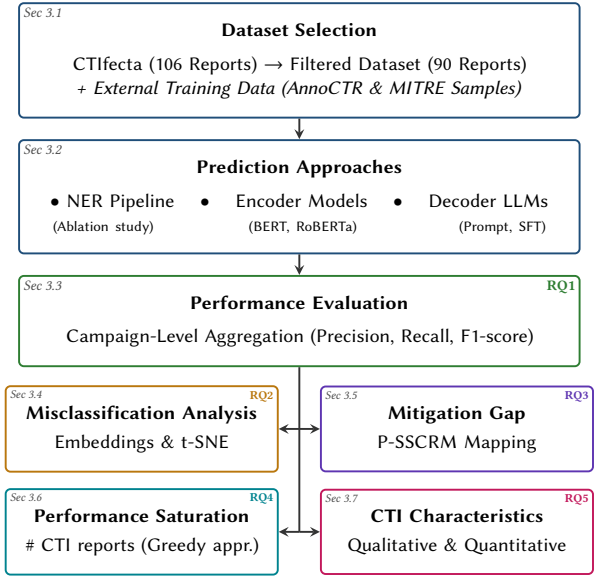
\begin{figure}[ht]
    \centering
    \resizebox{\linewidth}{!}{%
    \begin{tikzpicture}[
        >=stealth,
        font=\sffamily\small,
        databox/.style={draw=phasecolor, thick, align=center, minimum height=1cm, inner sep=8pt, rounded corners=2pt},
        predbox/.style={draw=phasecolor, thick, align=center, minimum height=1cm, inner sep=8pt, rounded corners=2pt},
        rq1box/.style={draw=rq1color, thick, align=center, minimum height=1cm, inner sep=8pt, rounded corners=2pt},
        rqbase/.style={thick, align=center, minimum height=1cm, inner sep=8pt, rounded corners=2pt, text width=3.4cm},
        rq2box/.style={rqbase, draw=rq2color},
        rq3box/.style={rqbase, draw=rq3color},
        rq4box/.style={rqbase, draw=rq4color},
        rq5box/.style={rqbase, draw=rq5color}
    ]

    \node[databox, text width=7.6cm] (data) {\textbf{Dataset Selection}\\[4pt] CTIfecta (106 Reports) $\rightarrow$ Filtered Dataset (90 Reports)\\ \textit{+ External Training Data (AnnoCTR \& MITRE Samples)}};
    \node[anchor=north west, font=\scriptsize\itshape\color{black!70}, inner sep=2pt] at (data.north west) {Sec \ref{sec:dataset_description}};

    \node[predbox, text width=7.6cm, below=0.3cm of data] (pred) {\textbf{Prediction Approaches}\\[4pt] $\bullet$ NER Pipeline \quad $\bullet$ \quad Encoder Models \quad $\bullet$ \quad Decoder LLMs\\ \scriptsize{(Ablation study)} \hspace{1.2cm} \scriptsize{(BERT, RoBERTa)} \hspace{1.3cm} \scriptsize{(Prompt, SFT)}};
    \node[anchor=north west, font=\scriptsize\itshape\color{black!70}, inner sep=2pt] at (pred.north west) {Sec \ref{sec:selected_methods}};

    \draw[->, thick, color=black!80] (data) -- (pred);

    \node[rq1box, text width=7.6cm, below=0.3cm of pred] (rq1) {\textbf{Performance Evaluation}\\[4pt] Campaign-Level Aggregation (Precision, Recall, F1-score)};
    \node[anchor=north west, font=\scriptsize\itshape\color{black!70}, inner sep=2pt] at (rq1.north west) {Sec \ref{sec:performance_metrics}};
    \node[anchor=north east, font=\scriptsize\bfseries\color{rq1color}, inner sep=2pt] at (rq1.north east) {RQ1};

    \draw[->, thick, color=black!80] (pred) -- (rq1);

    \coordinate (spine_top) at ([yshift=-0.9cm]rq1.south);
    \coordinate (spine_bot) at ([yshift=-2.4cm]rq1.south);
    \draw[thick, color=black!80] (rq1.south) -- (spine_bot);

    \node[rq2box, left=0.3cm of spine_top] (rq2) {\textbf{Misclassification Analysis}\\[2pt] Embeddings \& t-SNE};
    \node[anchor=north west, font=\tiny\itshape\color{black!70}, inner sep=2pt] at (rq2.north west) {Sec \ref{sec:factors}};
    \node[anchor=north east, font=\tiny\bfseries\color{rq2color}, inner sep=2pt] at (rq2.north east) {RQ2};

    \node[rq3box, right=0.3cm of spine_top] (rq3) {\textbf{Mitigation Gap}\\[2pt] P-SSCRM Mapping};
    \node[anchor=north west, font=\tiny\itshape\color{black!70}, inner sep=2pt] at (rq3.north west) {Sec \ref{sec:mitigation_controls}};
    \node[anchor=north east, font=\tiny\bfseries\color{rq3color}, inner sep=2pt] at (rq3.north east) {RQ3};

    \node[rq4box, left=0.3cm of spine_bot] (rq4) {\textbf{Performance Saturation}\\[2pt] \# CTI reports (Greedy appr.)};
    \node[anchor=north west, font=\tiny\itshape\color{black!70}, inner sep=2pt] at (rq4.north west) {Sec \ref{sec:performance_saturation}};
    \node[anchor=north east, font=\tiny\bfseries\color{rq4color}, inner sep=2pt] at (rq4.north east) {RQ4};

    \node[rq5box, right=0.3cm of spine_bot] (rq5) {\textbf{CTI Characteristics}\\[2pt] Qualitative \& Quantitative};
    \node[anchor=north west, font=\tiny\itshape\color{black!70}, inner sep=2pt] at (rq5.north west) {Sec \ref{sec:characteristics}};
    \node[anchor=north east, font=\tiny\bfseries\color{rq5color}, inner sep=2pt] at (rq5.north east) {RQ5};

    \draw[->, thick, color=black!80] (spine_top) -- (rq2.east);
    \draw[->, thick, color=black!80] (spine_top) -- (rq3.west);
    \draw[->, thick, color=black!80] (spine_bot) -- (rq4.east);
    \draw[->, thick, color=black!80] (spine_bot) -- (rq5.west);

    \end{tikzpicture}%
    }
    \caption{Overview of the experimental workflow.}
    \label{fig:study_design}
    \vspace{-4mm}
\end{figure}

\vspace{-2mm}
\section{Study Design}
\label{sec:methodology}
We employ a \textit{conceptual replication}~\cite{dennis2015replication} and \textit{extension} design~\cite{carver2010towards} to evaluate existing automated ATT\&CK technique extraction methods. A conceptual replication tests the same research phenomenon as prior work but in a different context. In this study, we evaluate the same 29 methods with settings across three approaches (\S~\ref{sec:selected_methods}) studied by Büchel et al.~\cite{buchel2025sok}, but move from a single-report evaluation to a campaign-level, multi-report dataset. An extension design builds on the replicated study by introducing additional research questions that go beyond the scope of the original. We extend the evaluation to analyze the causes of misclassification (\autoref{RQErrorAnalysis}), the downstream impact of controls (\autoref{RQMitigationGap}), performance saturation (\autoref{RQSaturation}), and report characteristics (\autoref{RQCTICharacteristics}), none of which were addressed in the original study. 

We provide an overview of our study design in \autoref{fig:study_design}. We first select a dataset of multiple CTI reports for the same campaign (\S\ref{sec:dataset_description}) and state-of-the-art methods under three approaches (\S\ref{sec:selected_methods}. Next, we evaluate their campaign-level performance (\S~\ref{sec:performance_metrics}). We then extract the misclassified and missed \textit{ATT\&CK} techniques and investigate the factors behind them(\S~\ref{sec:factors}) and quantify how those errors propagate into gaps in control coverage(\S~\ref{sec:mitigation_controls}). Finally, we examine how performance evolves as the number of CTI reports increases (\S~\ref{sec:performance_saturation}) and characterize the CTI reports that influence performance (\S~\ref{sec:characteristics}).

\vspace{-3mm}
\subsection{Dataset Selection}
\label{sec:dataset_description}
To evaluate our research questions, we separate our training and testing data into two distinct datasets. Initially, we considered the CTIfecta dataset by Hamer et al. \cite{hamer2026closing} for both purposes, as it provides multiple reports from the three attack campaigns required for our evaluation. However, more than 65\% (55 of 82) of its ATT\&CK techniques appear at most 5 times across the entire dataset, resulting in a long-tail distribution. Training directly on the long-tail CTIfecta data risks severe overfitting, bias toward majority classes, and poor generalization. Moreover, to the best of our knowledge, no other public campaign-level CTI dataset currently exists. Consequently, relying solely on CTIfecta for training would necessitate significant methodological interventions—such as transfer learning from domain-specific foundational models \cite{lin2022attack, aghaei2022securebert} or advanced data augmentation \cite{gao2023benefits, ruiz2025synthcti}. We therefore use the AnnoCTR dataset for training and fine-tuning (discussed in the next paragraph), reserving CTIfecta strictly for testing and evaluation.

\textbf{Training dataset.} For training or fine-tuning, we leverage the AnnoCTR dataset~\cite{lange2024annoctr}. Because MITRE ATT\&CK technique definitions are standardized and independent of campaign-specific context, the model learns relevant concepts while generalizing beyond specific campaigns. We selected AnnoCTR for its broader coverage (118 techniques) compared with alternatives such as TRAM (50 techniques) \cite{mitre_tram_2023}. However, 17 MITRE ATT\&CK techniques present in our testing data were not covered in AnnoCTR. To ensure complete label coverage, we collected additional samples for each missing technique directly from the MITRE ATT\&CK website, using the technique descriptions provided in the official ATT\&CK knowledge base. In the AnnoCTR dataset, minority classes occur 1-5 times, with a mean of 2.02 samples per class. We selected 3 samples per missing technique, consistent with the mean representation of minority classes in AnnoCTR, yielding 51 additional training samples in total.


\textbf{Testing dataset.} For testing, we use the CTIfecta dataset \cite{hamer2026closing}, comprising 106 CTI reports spanning three attack campaigns—SolarWinds, XZ Utils, and Log4j—with 30, 31, and 45 reports per campaign, respectively, authored by a range of organizations, including security vendors, government agencies, and incident response teams. For example, the Solarwinds attack campaign includes government directives (e.g., CISA ED21-01 (https://tinyurl.com/bdf9z568)
and in-depth industry forensic whitepapers (https://tinyurl.com/nz78pj2e). 
Together, these reports encompass 114 unique MITRE ATT\&CK techniques. We selected CTIfecta for its unique emphasis on \textit{depth} over \textit{breadth}. Unlike existing datasets (e.g., TRAM \cite{mitre_tram_2023}, AnnoCTR \cite{lange2024annoctr}, and TTPHunter \cite{rani2023ttphunter}) that prioritize broad, single-source coverage of disjoint attack campaigns, CTIfecta aggregates diverse reports describing the \textit{same} attack campaigns. This multi-view structure enables cross-report aggregation, capturing the semantic variability and reporting inconsistencies that broader single-report datasets miss. Since our objective is to automatically extract adversarial activities and map them to ATT\&CK techniques, we filter the dataset using inclusion and exclusion criteria. We retain only reports that contain at least one ATT\&CK technique and exclude reports that explicitly reference ATT\&CK technique identifiers in the text to avoid trivial mappings. After filtering, our final dataset comprises 90 CTI reports (21 SolarWinds, 28 XZ Utils, and 41 Log4j) that cover 82 unique attack techniques.

\begin{table*}[htbp]
\centering
\small
\caption{Summary of the 29 Evaluated Methods Across Three Extraction Approaches, with Model Architectures and Training Configurations Used in the Study.}
\label{tab:approaches_categories_methods}
\begin{tabular}{@{}p{1.8cm}p{2cm}p{5.5cm}p{6.2cm}@{}}
\toprule
\textbf{Approach (\#methods)} & \textbf{Category} & \textbf{Methods} & \textbf{Configuration \& Metadata} \\ \midrule
NER (6) & Rule-based &
full, base, no\_lemma, no\_parsing, no\_pos, no\_related\_words &
Ablation study; Syntactic/Semantic Pipeline (Rule-based, no training) \\ \midrule
\multirow{2}{1.8cm}{Encoder-based Classification (15)} &
BERT-based &
CySecBERT, SciBERT-\{c, uc\}, CyBERT, TRAM, SecureBERT, SecBERT, bert-base-\{c, uc\}, DarkBERT &
\multirow{2}{6.2cm}{\{c\}=cased, \{uc\}=uncased; \\ \{b\}=base, \{l\}=large; Multi-label classification head; Activation: Sigmoid; LR: $2{\times}10^{-5}$; Batch Size: 16} \\ \cmidrule(l){2-3}
 & RoBERTa-based &
SecRoBERTa, roberta-\{b, l\}, xlm-roberta-\{b, l\} & \\ \midrule
\multirow{2}{1.8cm}{Decoder-based LLM (8)} &
Prompt-based (Zero/Few-shot) &
RAW~\cite{sahoo2024systematic}, FSP, RAG, FSP+RAG &
\multirow{2}{6.2cm}{LLM: Llama-3.1-8B-Instruct~\cite{grattafiori2024llama}; RAG Embedding: Qwen2-7B-Instruct~\cite{li2023towards}; RAG Context: Top-5 retrieved techniques; PEFT: LoRA~\cite{hu2022lora} (16-bit); LR: $1{\times}10^{-5}$ (Emb), $2{\times}10^{-5}$; Training: 3 Epochs, Batch Size 4} \\ \cmidrule(l){2-3}
 & Weight-based (SFT) &
SFT-RAW, SFT-FSP, SFT-RAG, SFT-FSP+RAG & \\ \bottomrule
\end{tabular}
\vspace{-4mm}
\end{table*}

\vspace{-2mm}
\subsection{Selected Methods of Three Approaches}
\label{sec:selected_methods}
To evaluate existing approaches for mapping CTI reports of a campaign to ATT\&CK techniques, we replicate the 29 state-of-the-art methods evaluated by Büchel et al.~\cite{buchel2025sok}, spanning three approaches: named entity recognition (NER), encoder-based classification, and decoder-based LLM. We preserve the original configurations, hyperparameters, and model architectures as reported in~\cite{buchel2025sok}. A summary of all 29 methods, including their configurations, is presented in ~\autoref{tab:approaches_categories_methods}.

For \textbf{NER}, the full pipeline comprises five syntactic and semantic components. To assess the contribution of each component, we conduct an ablation study by systematically disabling one module at a time and measuring the resulting performance degradation relative to the full pipeline. Notably, the NER approach is rule- and pattern-driven and does not require model training. For \textbf{encoder-based classification}, 15 methods are evaluated, grouped into two categories based on model architecture: BERT-based and RoBERTa-based variants. For \textbf{decoder-based LLMs}, we evaluate eight methods categorized as: prompt-based and weight-based configurations. Within each category, four methods are considered: (i) zero-shot prompting using only the CTI report text (RAW), (ii) few-shot prompting with five randomly labeled examples (FSP), (iii) RAG supplying the top five most relevant ATT\&CK techniques based on cosine similarity of the given CTI text as context (RAG), and (iv) a combination of few-shot prompting and RAG (FSP+RAG).

\vspace{-2mm}
\subsection{Effectiveness of Extraction Methods}
\label{sec:performance_metrics}

We first obtain predictions from each automated method for every individual report. We then aggregate the predictions to the campaign level (e.g., SolarWinds) by taking the union of all \textit{ATT\&CK} techniques predicted across reports describing the same attack campaign. For example, the SolarWinds campaign comprises 21 CTI reports: if one method predicts \{\textit{T1078, T1027}\} for one report and \{\textit{T1027, T1195}\} for another report, the aggregated campaign-level prediction for these two reports is \{\textit{T1078, T1027, T1195}\}. We apply the same aggregation across all 21 reports in the SolarWinds campaign to obtain the full set of predicted campaign-level techniques. Finally, we evaluate the aggregated predictions against the complete set of ground-truth techniques annotated across all 21 reports in the SolarWinds campaign (\S~\ref{sec:dataset_description}) in the CTIfecta dataset.

We evaluate the methods' effectiveness using precision, recall, and F1 Score. \textit{Precision} measures the proportion of predicted techniques that are correct, \textit{recall} captures the proportion of ground-truth techniques successfully identified, and the \textit{F1-score} represents their harmonic mean. We report macro-averaged values, treating each technique equally regardless of its frequency in the dataset.

\vspace{-2mm}
\subsection{Misclassification and Missed ATT\&CK Techniques Analysis}
\label{sec:factors}
We identify missed and misclassified ATT\&CK techniques by quantitatively analyzing false negatives (FNs) and false positives (FPs) at the method level for each approach.  
We then investigate whether semantic similarity between ATT\&CK technique descriptions contributes to misclassifications and missed techniques. Techniques with similar official descriptions may be misclassified by models that rely on textual representations. 
To empirically examine whether a misclassification is attributable to semantic overlap in the ATT\&CK technique description, we compute text embeddings~\cite{reimers2019sentence} for the descriptions of all predicted and ground-truth techniques and then calculate cosine similarities~\cite{pedregosa2011scikit} for every possible pair of these techniques. 

Following prior work~\cite{cann2025using}, we apply a similarity threshold ($\delta \geq 0.7$) to identify similar pairs and report the percentage of FPs and FNs whose descriptions fall above the threshold. Finally, to visually represent the relationships, we project the embeddings into a two-dimensional space using t-SNE~\cite{van2008visualizing}, where points that are close together indicate higher semantic similarity between their technique descriptions.

\vspace{-2mm}
\subsection{ATT\&CK Technique to Control Mapping}
\label{sec:mitigation_controls}
Some organizations want to understand attacker technique trends so they can prioritize the adoption of controls to protect against them. Within the CTIfecta dataset, each attack campaign is mapped to a ground-truth set of ATT\&CK techniques and their corresponding controls. If an automated method fails to extract a required ATT\&CK technique, the corresponding controls may also be missed, leaving the organization exposed to undetected vulnerabilities. To assess the impact of extraction errors, we map each predicted and ground-truth \textit{ATT\&CK} technique to its corresponding Proactive Software Supply Chain Risk Management (P-SSCRM) controls~\cite{williams2024proactive}. The P-SSCRM framework unifies 73 controls (referred to as tasks) across 10 government and industry standards. We use P-SSCRM rather than native MITRE mitigations because it aligns with the standards practitioners consult and provides broader cross-standard coverage with actionable guidance.

\textbf{Baseline Mapping.} We use the \textit{ATT\&CK} technique to P-SSCRM mapping data published by Hamer et al.~\cite{hamer2026closing} as our baseline. Their original dataset contains 4,453 candidate mappings of ATT\&CK techniques to control pairs spanning 198 attack techniques. Of these, 97 techniques were mapped using triangulation across four distinct mapping strategies. We considered these 97 techniques as confirmed baseline mapping.

\textbf{Extended Manual Mapping Protocol.} The above mapping dataset also provides individual outputs for each strategy across all techniques. We leverage both the confirmed mappings and the available strategy outputs to support our mapping process. Our evaluation dataset contains 82 unique \textit{ATT\&CK} techniques (\S~\ref{sec:dataset_description}), of which 44 are covered by the confirmed baseline mappings. The remaining 38 out of 82 techniques, therefore, require additional mapping for our analysis. In addition, we found 6 \textit{ATT\&CK} techniques from the automated methods' prediction that are not covered by the confirmed baseline mappings. Consequently, 38+6=44 techniques were manually mapped to extend the baseline dataset for this study.
For the 44 techniques not covered by the confirmed baseline mappings, we analyzed 437 candidate technique–to–control pairs available in the baseline dataset. We performed the mapping in two phases. In the first phase, we applied a filtering criterion to prioritize higher-confidence candidate pairs: a pair was retained only if it achieved agreement between the manual review strategy and at least one of the three automated mapping strategies reported in the baseline dataset. In the second phase, two co-authors independently evaluated the remaining pairs by cross-referencing the \textit{MITRE ATT\&CK} technique descriptions and mitigation strategies against the objectives, descriptions, and assessment questions of the corresponding P-SSCRM controls. This independent review process resulted in 26 inter-rater disagreements, which were resolved by the third author. 

\textbf{Impact Measurement.} With the complete mapping established, we derive two sets of controls for each attack campaign: one from the ground-truth \textit{ATT\&CK} techniques annotated in the CTI reports, and another one from the \textit{ATT\&CK} techniques predicted by each automated method. Inspired by Hamer et al. \cite{hamer2026closing}, we use metrics to calculate the errors of automated misclassifications on mapped controls. First, for each attack campaign, we calculate if each control is: \Circled{1} \textbf{\texttt{Matched controls:}} Controls present in both the ground truth and predicted sets. \Circled{2} \textbf{\texttt{Missed controls:}} Ground truth controls absent from the predicted set, resulting from missed attack techniques. \Circled{3} \textbf{\texttt{Unnecessary controls:}} Invalid predicted controls absent from the ground truth set, resulting from wrong attack technique predictions.
Finally, we calculate \textit{control coverage} as the percentage of \texttt{matched controls} divided by the sum of \texttt{matched} and \texttt{unnecessary} controls.
\vspace{-4mm}
\subsection{Performance Saturation Analysis}
\label{sec:performance_saturation}
Since no single CTI report captures the full scope of an attack campaign, aggregating reports is essential, but it raises the practical question of how many are needed before information extraction saturates. To identify the minimum number of reports required for performance saturation, we perform a greedy, incremental evaluation for each attack campaign, adding the most informative reports first. Inspired by the concept of \textit{code saturation~\cite{hennink2017code}} (the point at which no new techniques have been identified), we track the discovery of techniques as more reports are incorporated. For each campaign, we rank CTI reports by their true-positive contribution, defined as the number of techniques each report correctly identifies. If multiple reports contribute the same number of techniques, we resolve the tie by ordering them alphabetically by their file names. We then incrementally add reports in descending order of contribution. After each addition, we calculate the cumulative evaluation metrics (precision, recall, and F1-score) against a fixed ground-truth reference set containing all attack techniques for that campaign (\S~\ref{sec:performance_metrics}).

To identify when additional reports provide limited benefit, prior work~\cite{guest2020simple} recommends $\leq$0.05, and we adopt it to define performance saturation as the point at which improvements in evaluation metrics fall below this threshold. The number of reports at this point represents the minimum required to achieve maximum performance.


\subsection{CTI Report Characteristic Analysis}
\label{sec:characteristics}
To examine which characteristics of CTI reports influence the automated extraction of ATT\&CK techniques, we select the best-performing method from each of the three approaches based on the highest campaign-level recall (\S~\ref{sec:performance_metrics}). We hypothesize that reports that include more ATT\&CK techniques provide more descriptions of a campaign. We then group reports for each campaign based on the recall saturation point (identified in \S~\ref{sec:performance_saturation}). Reports that appear prior to the saturation point and contribute the majority of the true positives are designated as \textit{pre-segment} reports, while reports added after the saturation point are designated as \textit{post-segment} reports. 

Then, for each CTI report in the two groups, we extract the following features as characteristics to analyze which characteristics influence the automated methods to prioritize reports in the pre-segment:
\Circled{1} \textbf{Word Count:} The total number of words in a report.
\Circled{2} \textbf{Sentence Count:} The total number of sentences in a report.
\Circled{3} \textbf{Readability Score:} The Flesch Reading Ease score~\cite{kincaid1975derivation}, ranging from 1 to 100, where higher values indicate greater readability.
\Circled{4} \textbf{Vendor:} The publishing organization of the report. We hypothesize that some vendors document CTI reports better than others.
\Circled{5} \textbf{Publication date:} The publication date of a report. 

Finally, we statistically compare pre-saturation and post-saturation reports to determine whether their characteristics differ significantly. For all numeric and ordinal metrics, we apply the Mann–Whitney U test~\cite{macfarland2016mann} due to the non-normal distribution of report characteristics and unequal group sizes. To quantify the magnitude of differences, we report Cliff’s delta ($\delta$) as a non-parametric effect size measure. We exclude the vendor from the statistical test because it is a nominal categorical variable with no inherent ordering. Instead, we analyze vendor distributions descriptively by comparing vendor frequencies across pre- and post-saturation report sets.

\begin{figure}
    \centering
    \includegraphics[width=\linewidth]{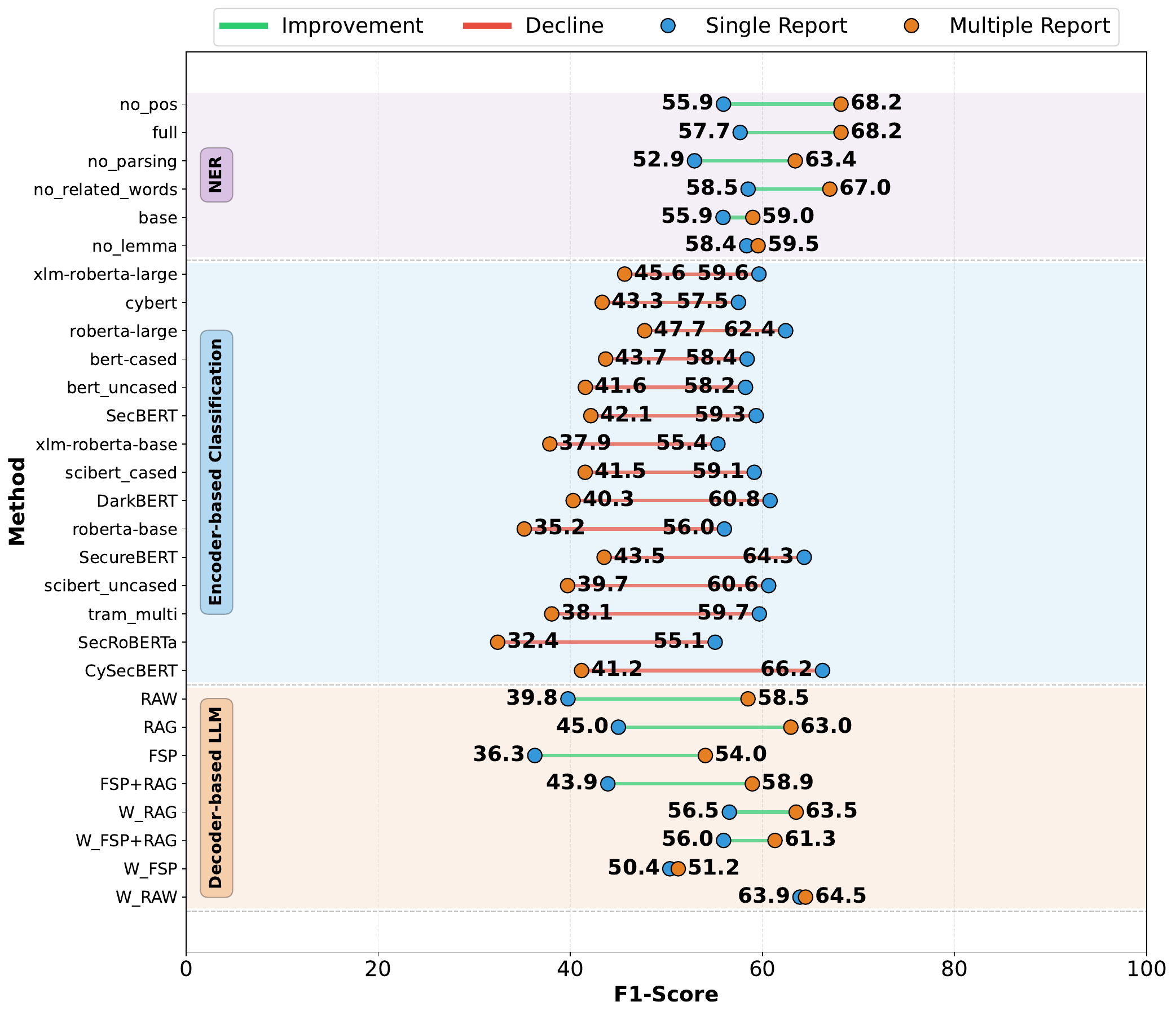}
    \caption{Impact of multi-report on ATT\&CK technique extraction. Each dumbbell represents a method's performance shift from \textit{Single Report} ({\color{blue}$\bullet$}) to \textit{Multiple Reports} ({\color{orange}$\bullet$}). Green lines(\textcolor{green!60!black}{---}) indicate an improvement with multiple reports, while red lines (\textcolor{red!60!black}{---}) indicate a decline. }
    \label{fig:effectiveness}
    \vspace{-6mm}
\end{figure}
\section{Results}
\label{sec:results}

\subsection*{\reuseRQ{RQEffectiveness} \RQEffectiveness}
To answer \autoref{RQEffectiveness}, we replicate 29 methods from the three approaches mentioned in \autoref{tab:approaches_categories_methods} in Buchel et al.~\cite{buchel2025sok} on the campaign with multiple reports and compare them with the prior single report with respect to precision, recall, and F1-score. As we have three campaigns and prior studies have two datasets of single reports, we are averaging the three campaigns' performance and the two single reports' performance, and then comparing the performance between single- and multiple-report campaigns. To determine the performance difference, we performed the Mann-Whitney U test~\cite{macfarland2016mann}, and to measure the effect, we performed Cliff's delta ($\Delta$). We compared the performance of methods within each approach using the median, as some methods exhibit skewed performance distributions, making the median a more robust measure to minimize the influence of outliers. The results of the F1-score are shown in \autoref{fig:effectiveness}, and precision and recall are provided in the replication package. 

All eight \textit{decoder-based LLM} methods demonstrate consistent performance improvements on the multi-report campaign dataset. The median F1-score increases from 47.67 to 60.12, corresponding to a relative improvement of 26.1\%. This gain is accompanied by improvements in precision (66.80 $\rightarrow$ 72.10, +7.9\%) and, more notably, recall (43.98 $\rightarrow$ 56.50, +28.5\%). The improvement is statistically significant ($p \leq 0.001$) with a large effect size (Cliff's $\Delta$ = +0.656), indicating an advantage of the multi-report setting.

All six \textit{NER} methods also demonstrate consistent improvements on the multi-report campaign dataset, although the magnitude of improvement is more moderate compared to the \textit{decoder-based LLM} approach. The median F1-score increases from 56.80 to 65.21, corresponding to a relative improvement of 14.8\%. This improvement is accompanied by gains in precision (55.90 $\rightarrow$ 59.00, +5.5\%) and recall (60.00 $\rightarrow$ 64.40, +7.3\%). The observed improvements are associated with very large effect sizes, with Cliff's $\Delta$ = +1.00, indicating a strong and consistent advantage of the multi-report setting. 

In contrast to \textit{decoder-based LLM} and \textit{NER} approaches, all fifteen \textit{encoder-based classification} methods demonstrate consistent performance degradation on the multi-report campaign dataset. The median F1-score decreases from 59.34 to 41.54, corresponding to a relative decline of 30.0\%. This decline is accompanied by decreases in precision (58.50 $\rightarrow$ 56.10, -4.1\%) and, more severely, recall (61.90 $\rightarrow$ 39.68, -35.9\%). The decline is statistically significant ($p \leq 0.001$) with a large negative effect size (Cliff's $\Delta$ = -1.00).

\begin{boxK}
\vspace{-2mm}
\textit{\autoref{RQEffectiveness}:} Overall, our results show that \textit{NER} and \textit{decoder-based LLM} approaches achieve better performance on multi-reports, whereas the \textit{encoder-based classification} approach achieves better performance on a single report. No single method dominates uniformly across all three campaigns.
\vspace{-2mm}
\end{boxK}

\begin{figure}[ht]
    \centering
    \begin{subfigure}
        [b]{0.48\linewidth}
        \centering
        \includegraphics[width=\linewidth]{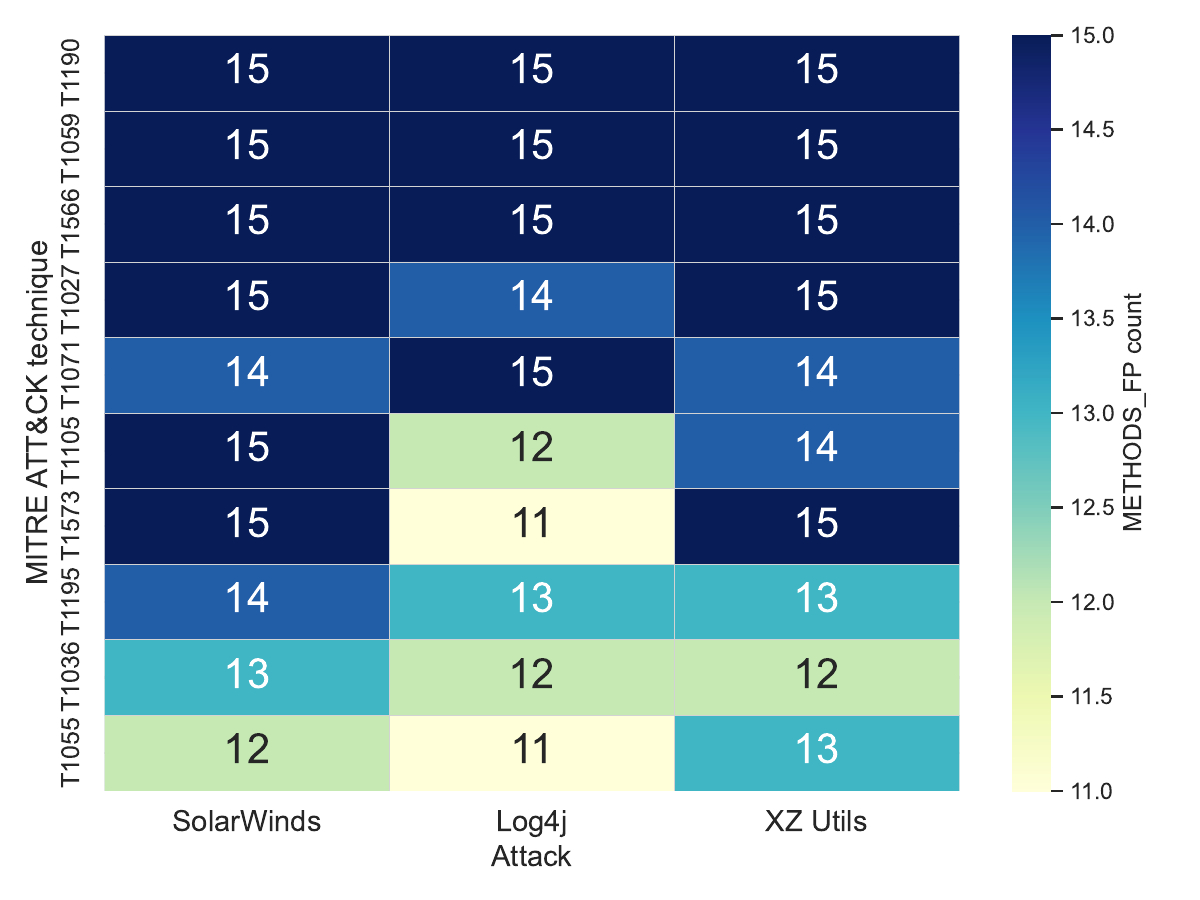}
        \caption{Classification (15 methods)}
    \end{subfigure}
    \hfill
    \begin{subfigure}
        [b]{0.48\linewidth}
        \centering
        \includegraphics[width=\linewidth]{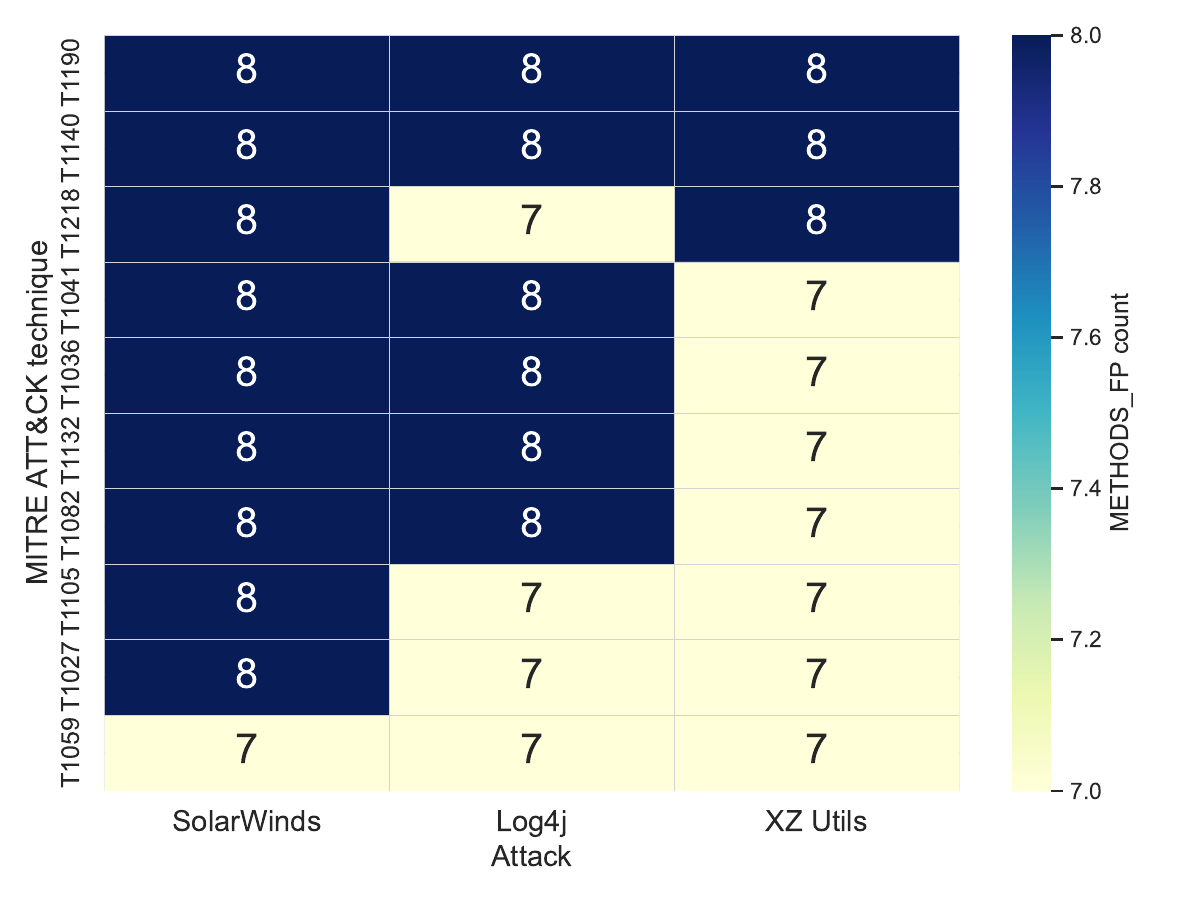}
        \caption{LLM (8 methods)}
    \end{subfigure}
    \caption{Comparison of \textbf{false positive (FP)} across different approaches of Solarwinds, XZ Utils, and Log4j attacks. Rows correspond to specific MITRE ATT\&CK techniques, columns represent the attack, and cell values indicate the number of methods incorrectly predicting that technique. There are no FPs for the NER approach.}
    \label{fig:FP_analysis}
    \vspace{-4mm}
\end{figure}

\begin{figure*}[ht]
    \centering
    \begin{subfigure}
        [b]{0.33\textwidth}
        \centering
        \includegraphics[width=\textwidth]{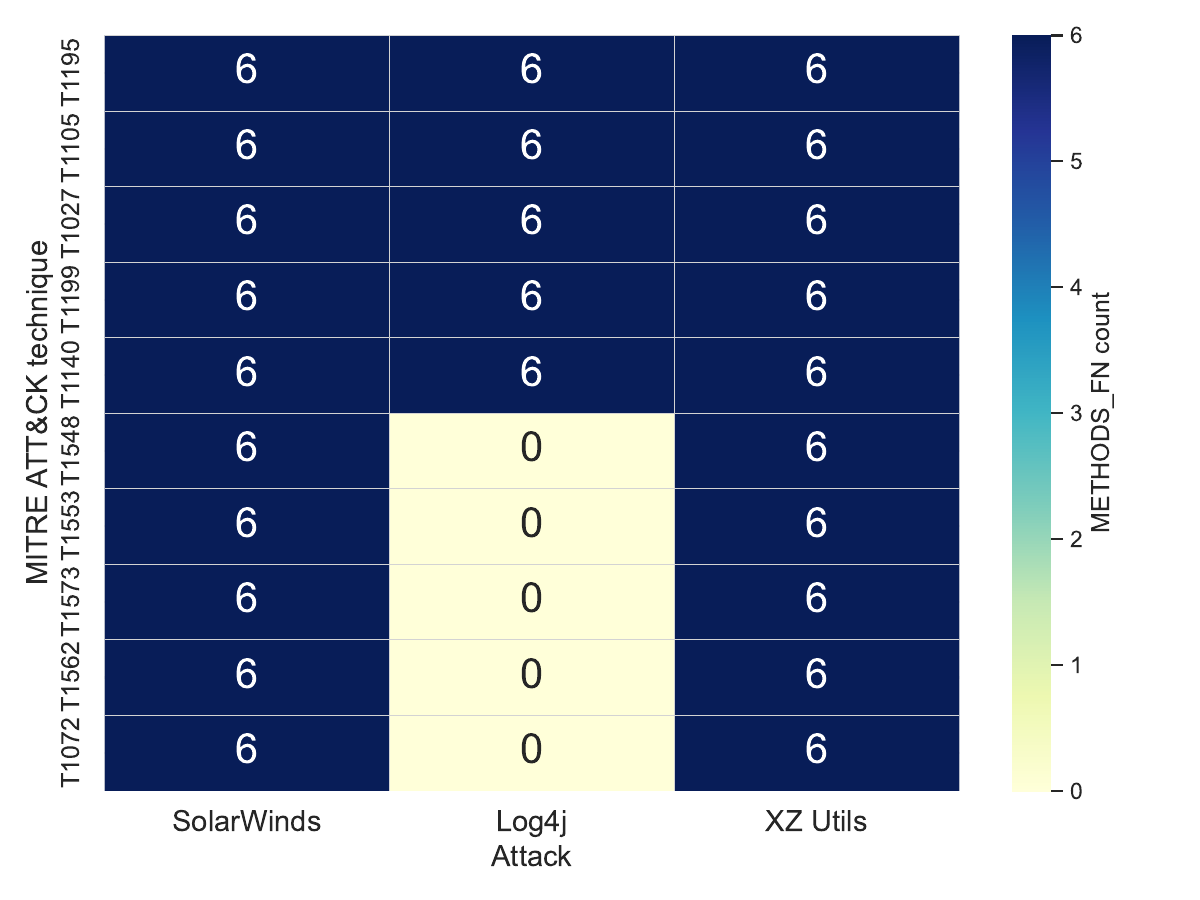}
        \caption{NER (6 methods)}
    \end{subfigure}
    \hfill
    \begin{subfigure}
        [b]{0.33\textwidth}
        \centering
        \includegraphics[width=\textwidth]{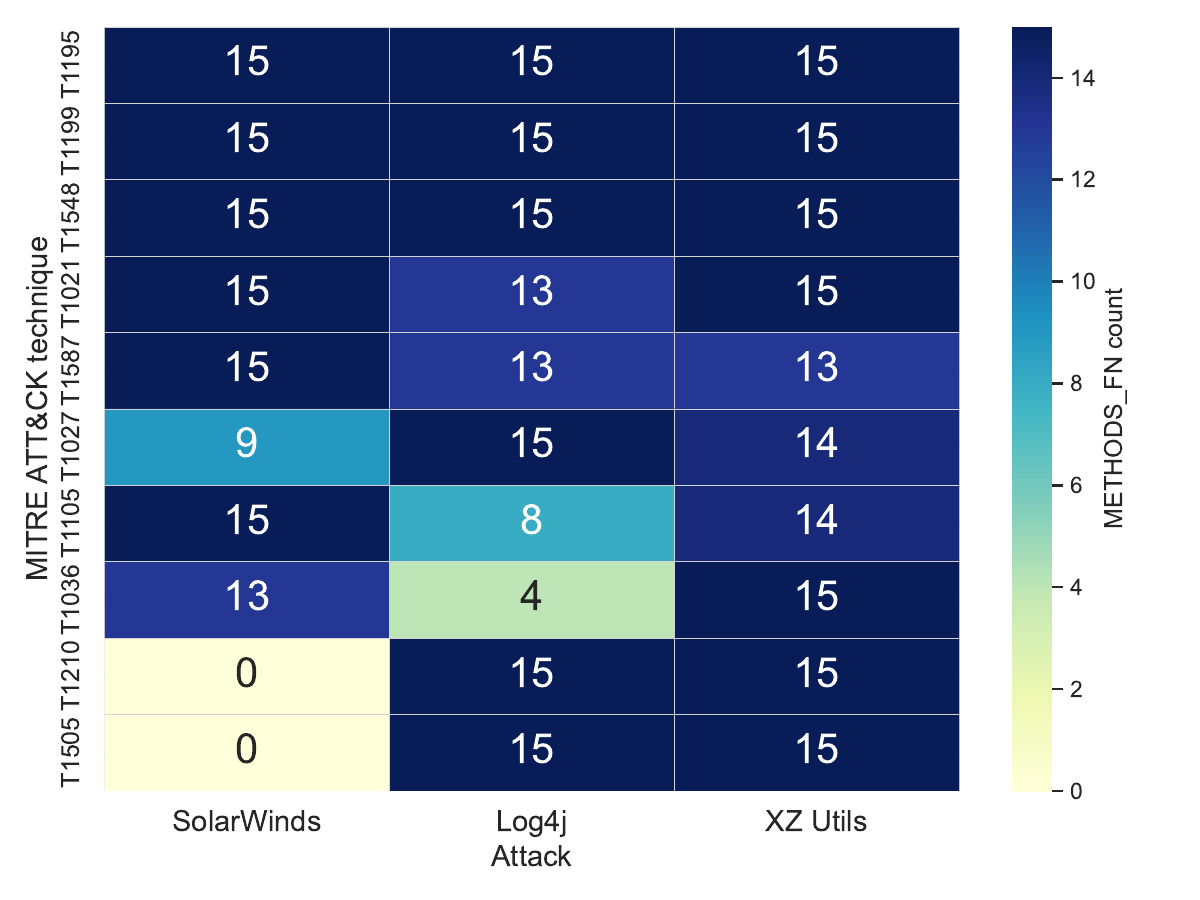}
        \caption{Encoder-based classification (15 meth.)}
    \end{subfigure}
    \hfill
    \begin{subfigure}
        [b]{0.33\textwidth}
        \centering
        \includegraphics[width=\textwidth]{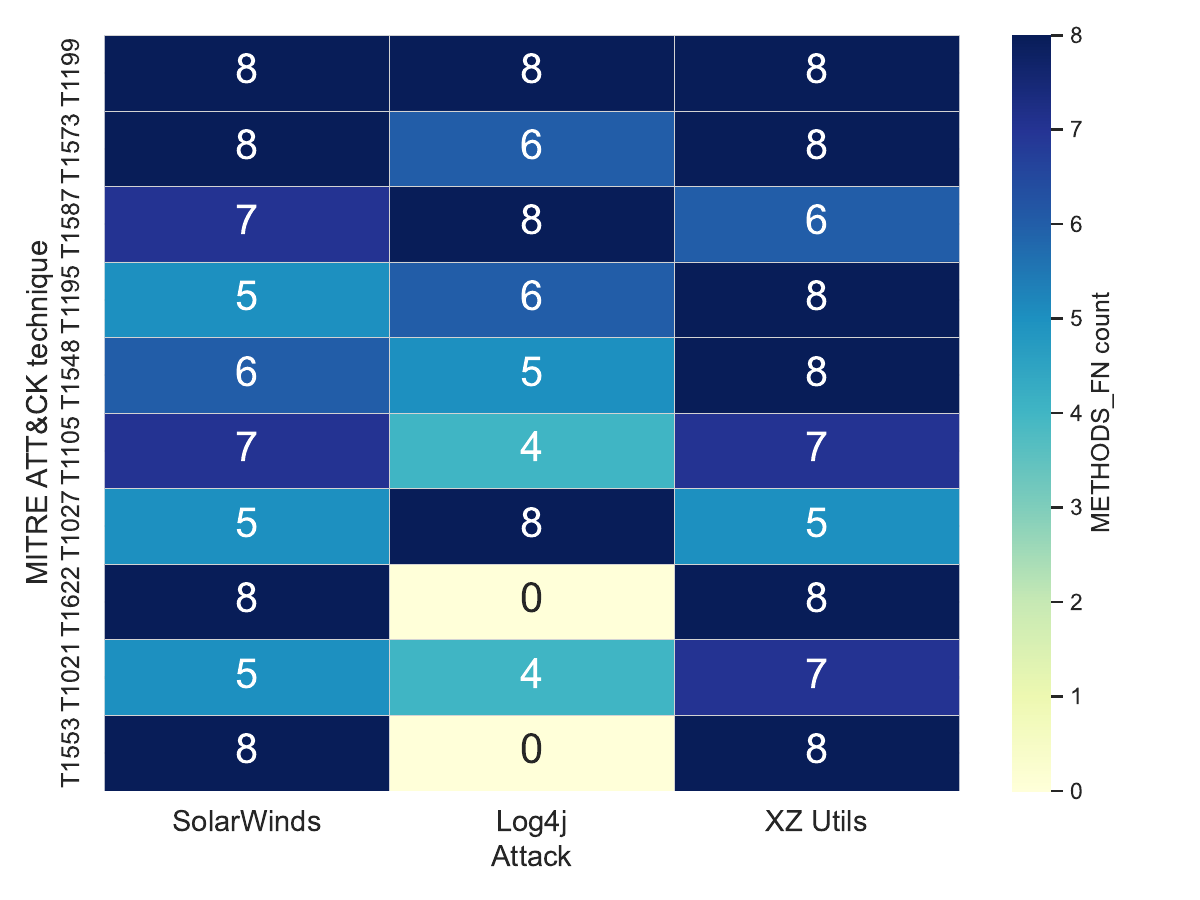}
        \caption{Decoder-based LLM (8 methods)}
    \end{subfigure}
    \caption{Comparison of \textbf{false negative} across different approaches of Solarwinds, Log4j, and XZ Utils. Rows correspond to techniques, columns represent campaign, and cell values indicate \#methods failing to detect that technique.}
    \label{fig:FN_analysis}
    \vspace{-4mm}
\end{figure*}

\begin{figure}
    \centering
    \includegraphics[width=0.90\linewidth]{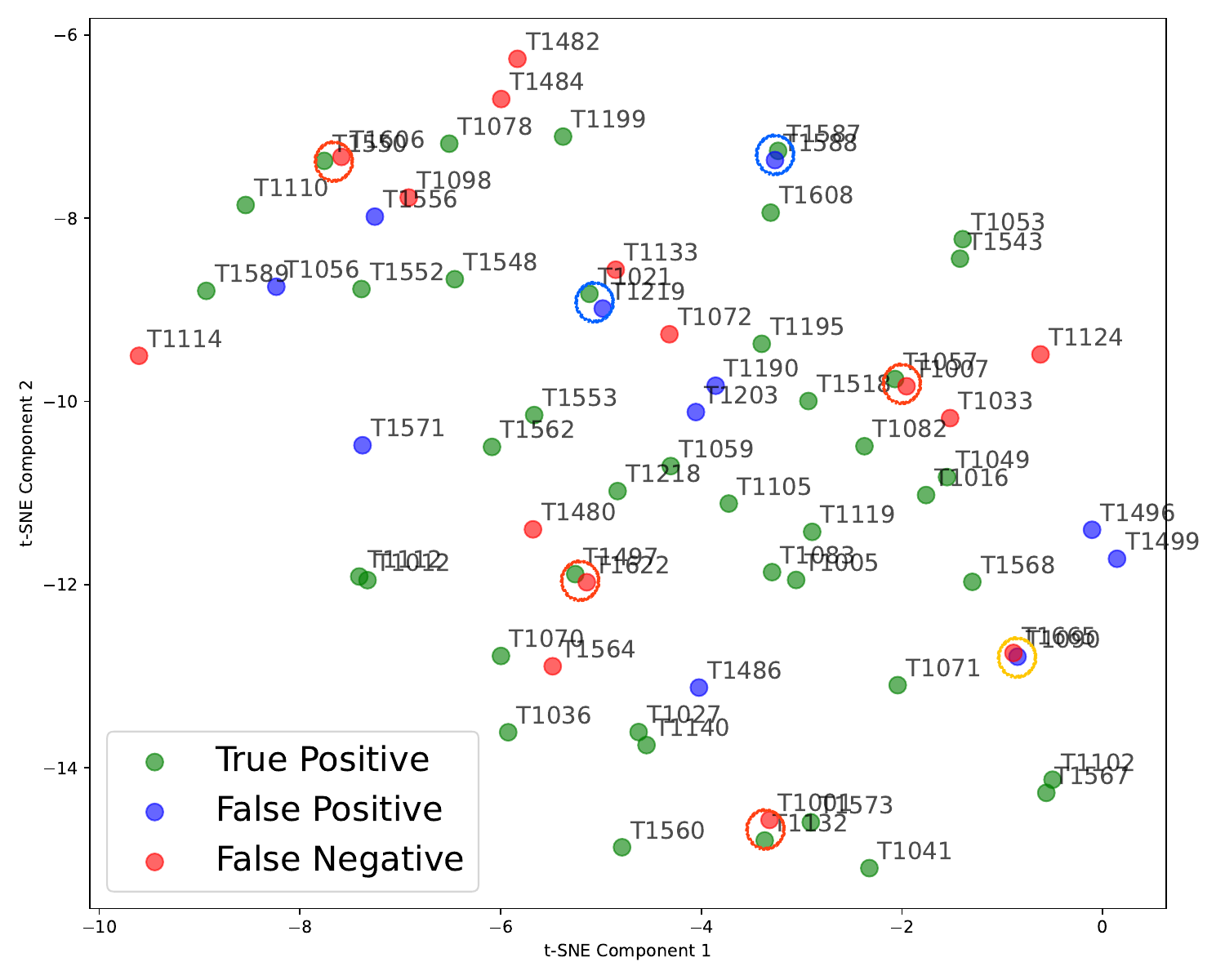}
    \caption{T-SNE projection of ATT\&CK technique descriptions, illustrating the semantic similarity between ground-truth techniques from the SolarWinds campaign and predictions from the SFT-RAG method.}
    \label{fig:sft_raw_solarwinds_t-sne}
    \vspace{-4mm}
\end{figure}

\subsection*{\reuseRQ{RQErrorAnalysis} \RQErrorAnalysis}
To answer \autoref{RQErrorAnalysis}, we evaluate false positives (FPs) for misclassified techniques and false negatives (FNs) for missed techniques. Figures \ref{fig:FP_analysis} and \ref{fig:FN_analysis} present the top 10 misclassified and missed attack techniques across SolarWinds, Log4J, and XZ Utils (detailed counts are available in our replication package). Building on the embedding analysis described in \S~\ref{sec:factors}, we examine these errors to determine how semantic overlap within the MITRE ATT\&CK technique description drives misclassification.

\textbf{False Positive (FP) analysis:} From \autoref{fig:FP_analysis}, we observe that most existing methods consistently misclassify certain attack techniques. For example, \emph{T1190: Exploit Public-Facing Application}, \emph{T1059: Command and Scripting Interpreter}, and \emph{T1566: Phishing} are wrongly predicted by all 15 encoder-based classification methods, while \emph{T1190} and \emph{T1140: Deobfuscate/Decode Files or Information} are misclassified by all 8 decoder-based LLM methods. To understand what drives these consistent errors, we quantitatively analyzed the best-performing method (SFT RAG). Applying our cosine similarity threshold ($\ge$ 0.7) reveals that a substantial percentage of FPs share high semantic similarity with the actual ground truth: $33.3\%$ ($5/15$) for SolarWinds, $19.2\%$ ($5/26$) for XZ Utils, and $12.5\%$ ($3/24$) for Log4j. In the t-SNE projection (\autoref{fig:sft_raw_solarwinds_t-sne}), these highly similar pairs correspond to the highlighted clusters (\textcolor{blue}{dotted blue circled pairs}). For example, \emph{T1587: Develop Capabilities} is confused with \emph{T1588: Obtain Capabilities}, as both describe preparatory adversarial behaviors. Similarly, \emph{T1219: Remote Access Software} maps closely to \emph{T1021: Remote Services}. Furthermore, these misclassifications frequently occur within the same tactical boundaries. Across SolarWinds, XZ Utils, and Log4j, 40.0\% (6/15), 50.0\% (13/26), and 79.2\% (19/24) of FPs, respectively, share at least one tactic with their nearest ground-truth technique, with \emph{Defense Evasion} and \emph{Discovery} being the most frequently shared tactics across all three attacks. 

\textbf{False Negative (FN) analysis:} We observe a similar pattern for FNs, where specific ground-truth techniques are consistently missed by existing methods, as shown in \autoref{fig:FN_analysis}. To understand the drivers behind these missed techniques, we similarly analyze the best-performing method (\texttt{SFT-RAG}) with a cosine similarity threshold ($\ge$ 0.7) and reveal that a notable portion of missed techniques exhibit high semantic similarity to the incorrectly predicted labels: 20.0\% (2/10) for SolarWinds, 6.7\% (1/15) for XZ Utils, and 33.3\% (2/6) for Log4j. In the t-SNE projection (\autoref{fig:sft_raw_solarwinds_t-sne}), these missed ground-truth techniques closely cluster near the predicted ones (\textcolor{red}{dotted red circled pairs}). For instance, \emph{T1550: Use Alternate Authentication Material} is often missed in favor of \emph{T1606: Forge Web Credentials}. Additionally, we observed instances (\textcolor{yellow!80!black}{dotted yellow circled pair}) where an incorrect prediction maps closely to an FN, such as the confusion between \emph{T1090} (FP) and \emph{T1665} (FN). Furthermore, 70.0\% (7/10), 73.3\% (11/15), and 33.3\% (2/6) of FNs in SolarWinds, XZ Utils, and Log4J, respectively, share at least one tactic with their nearest predicted technique, a higher rate than observed for FPs, with \emph{Defense Evasion}, \emph{Discovery}, and \emph{Persistence} being the most frequently shared tactics across all three attacks.

\begin{boxK}
\vspace{-2mm}
\textit{\autoref{RQErrorAnalysis}:} Both misclassified and missed techniques cluster near their ground-truth counterparts in the embedding space, accounting for up to 33.3\% of FPs and FNs. These errors are further compounded by shared tactics, with up to 79.2\% of FPs and 73.3\% of FNs sharing at least one MITRE tactic with the nearest ground-truth technique.
\vspace{-2mm}
\end{boxK}


\subsection*{\reuseRQ{RQMitigationGap} \RQMitigationGap}

\begin{table}[ht]
\centering
\caption{Missed controls and mitigation coverage by approach. Format: \textit{Missed Controls (Coverage\%)}. Higher coverage indicates better performance. Total ground-truth controls are shown in column headers.}
\label{tab:missingControls}
\footnotesize
\setlength{\tabcolsep}{3pt} 
\resizebox{\columnwidth}{!}{%
\begin{tabular}{@{}llccc@{}}
\toprule
\textbf{Approach} & \textbf{Method} & \textbf{SolarWinds (35)} & \textbf{XZ Utils (32)} & \textbf{Log4j (38)} \\
\midrule
\multirow{6}{*}{NER} & base & 20 (42.9\%) & 23 (28.1\%) & 36 (5.3\%) \\
 & \bestcell{full} & \bestcell{19 (45.7\%)} & \bestcell{21 (34.4\%)} & \bestcell{33 (13.2\%)} \\
 & no\_lemma & 20 (42.9\%) & 23 (28.1\%) & 36 (5.3\%) \\
 & no\_parsing & 19 (45.7\%) & 21 (34.4\%) & 36 (5.3\%) \\
 & no\_pos & 19 (45.7\%) & 21 (34.4\%) & 33 (13.2\%) \\
 & no\_related\_words & 19 (45.7\%) & 21 (34.4\%) & 33 (13.2\%) \\
\midrule
\multirow{15}{*}{\shortstack[l]{Encoder-based \\ classification}} & CySecBERT & 23 (34.3\%) & 24 (25.0\%) & 22 (42.1\%) \\
 & DarkBERT & 25 (28.6\%) & 24 (25.0\%) & 22 (42.1\%) \\
 & SecBERT & 24 (31.4\%) & 24 (25.0\%) & 22 (42.1\%) \\
 & SecRoBERTa & 25 (28.6\%) & 32 (0.0\%) & 32 (15.8\%) \\
 & SecureBERT & 23 (34.3\%) & 24 (25.0\%) & 22 (42.1\%) \\
 & bert-base-cased & 22 (37.1\%) & 24 (25.0\%) & 20 (47.4\%) \\
 & bert-base-uncased & 24 (31.4\%) & 24 (25.0\%) & 22 (42.1\%) \\
 & cybert & 23 (34.3\%) & 24 (25.0\%) & 21 (44.7\%) \\
 & roberta-base & 25 (28.6\%) & 24 (25.0\%) & 23 (39.5\%) \\
 & \bestcell{roberta-large} & \bestcell{22 (37.1\%)} & \bestcell{24 (25.0\%)} & \bestcell{22 (42.1\%)} \\
 & scibert\_scivocab\_cased & 25 (28.6\%) & 24 (25.0\%) & 22 (42.1\%) \\
 & scibert\_scivocab\_uncased & 25 (28.6\%) & 24 (25.0\%) & 22 (42.1\%) \\
 & tram\_multi\_label\_model & 25 (28.6\%) & 24 (25.0\%) & 22 (42.1\%) \\
 & xlm-roberta-base & 25 (28.6\%) & 24 (25.0\%) & 22 (42.1\%) \\
 & xlm-roberta-large & 24 (31.4\%) & 24 (25.0\%) & 22 (42.1\%) \\
\midrule
\multirow{8}{*}{\shortstack[l]{Decoder-based \\ LLM}} & prompt\_FSP & 23 (34.3\%) & 22 (31.3\%) & 19 (50.0\%) \\
 & prompt\_RAG + FSP & 10 (71.4\%) & 18 (43.8\%) & 15 (60.5\%) \\
 & \textbf{\bestcell{prompt\_RAG}} & \textbf{\bestcell{8 (77.1\%)}} & \textbf{\bestcell{8 (75.0\%)}} & \textbf{\bestcell{11 (71.1\%)}} \\
 & prompt\_Raw & 11 (68.6\%) & 21 (34.4\%) & 11 (71.1\%) \\
 & weight\_SFT Raw & 23 (34.3\%) & 24 (25.0\%) & 23 (39.5\%) \\
 & weight+prompt\_SFT FSP & 15 (57.1\%) & 20 (37.5\%) & 14 (63.2\%) \\
 & weight+prompt\_SFT RAG + FSP & 8 (77.1\%) & 17 (46.9\%) & 9 (76.3\%) \\
 & weight+prompt\_SFT RAG & 14 (60.0\%) & 23 (28.1\%) & 8 (78.9\%) \\
\bottomrule
\end{tabular}%
}
\vspace{-6mm}
\end{table}

To answer \autoref{RQMitigationGap}, we first analyze the quantitative performance of mitigation control coverage across the three automated extraction approaches on the SolarWinds, XZ Utils, and Log4j datasets. We then qualitatively group the persistently missed controls into the P-SSCRM framework groups to identify systemic extraction gaps. \autoref{tab:missingControls} presents the number of missed controls and mitigation coverage for each method across the three attack campaigns. The names of the missed mitigation controls are provided in the replication package. 

Overall, decoder-based LLM methods achieve the highest mitigation coverage (77.1\%) across all attack campaigns, whereas NER and encoder-based classification methods miss the majority of ground-truth controls.
For example, the best NER method (\texttt{full}) achieves 45.7\% coverage on SolarWinds but drops to 13.2\% on Log4j, missing 33 of 38 ground-truth controls. Similarly, the 15 evaluated encoder-based methods show limited effectiveness, with coverage ranging from 0.0\% to 47.4\% (\texttt{bert-base-cased} on Log4j). In contrast, decoder-based LLM substantially improves coverage, particularly when combined with RAG. The \texttt{prompt\_RAG} method performs best, achieving 77.1\% coverage on SolarWinds and 75.0\% on XZ Utils (both missing 8 controls), while maintaining 71.1\% coverage on Log4j (11 missed controls).

Despite the relative success of decoder-based LLMs compared to NER and classification, failures to map certain controls reveal critical gaps in the Governance (G), Deployment (D), and Environment (E) groups of the P-SSCRM framework. Even the best-performing \texttt{prompt\_RAG} method misses controls in the G, D, and E groups. For example, operational environment controls (E.3.3, E.3.4, E.3.6, E.3.7), intrusion monitoring (D.2.1), and supplier management (G.3.4) are consistently omitted by all 29 evaluated methods across the three attack campaigns. 

\begin{boxK}
\vspace{-2mm}
\textit{\autoref{RQMitigationGap}:} While decoder-based LLMs with RAG significantly outperform NER and encoder-based methods, achieving up to 77.1\% mitigation coverage, all 29 evaluated methods exhibit systemic gaps in the Governance (G), Deployment (D), and Environment (E) domains of the P-SSCRM framework.
\vspace{-3.5mm}
\end{boxK}

\begin{figure*}
    \centering
    \includegraphics[width=0.9\textwidth]{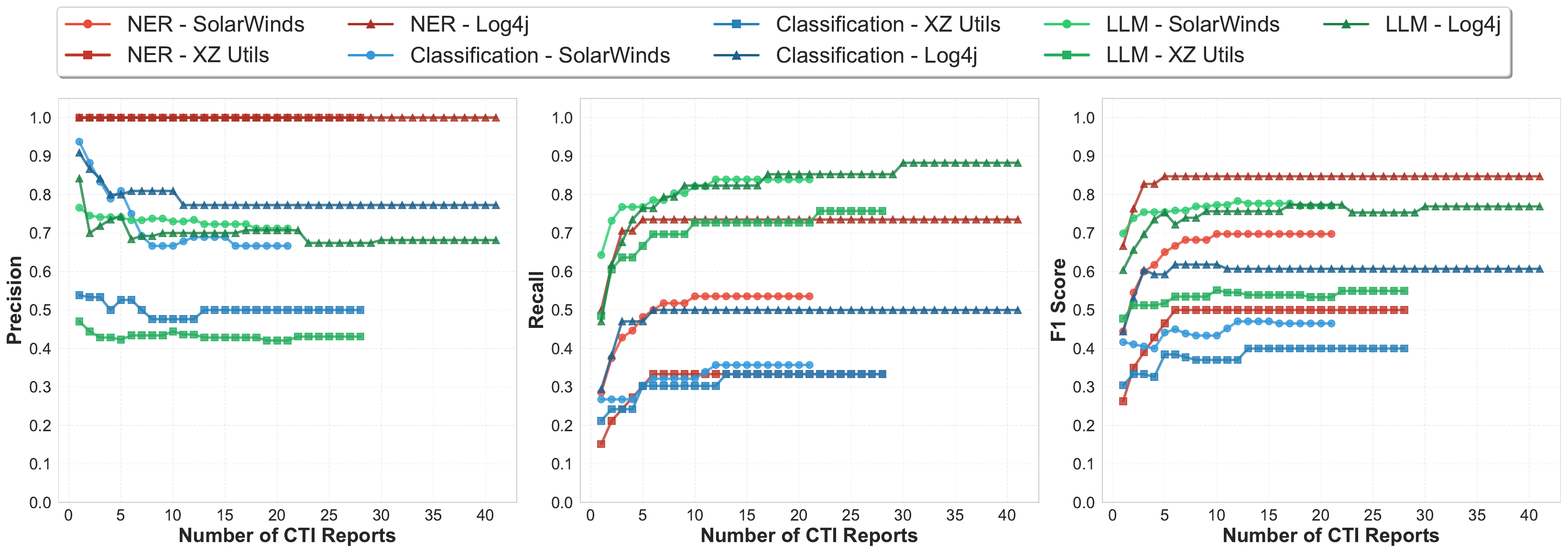}
    \caption{Performance comparison of NER, Encoder-based classification, and decoder-based LLM approaches across different number of CTI reports. The distinct lines, distinguished by color and marker style, represent the combinations of the three approaches evaluated across the SolarWinds, XZ Utils, and Log4j attack campaigns.}
    \label{fig:performane_vs_cti}
    \vspace{-2mm}
\end{figure*}

\vspace{-2mm}
\subsection*{\reuseRQ{RQSaturation} \RQSaturation}
To answer \autoref{RQSaturation}, we used a greedy incremental evaluation strategy described in \S~\ref{sec:performance_saturation}. \autoref{fig:performane_vs_cti} shows cumulative performance (precision, recall, F1) as additional CTI reports are incorporated, following the incremental evaluation procedure described in ~\autoref{sec:performance_saturation}.



For \textbf{Precision}, NER reaches saturation after a single CTI report across all three attack campaigns, achieving maximum precision (1.0). In contrast, encoder-based classification and decoder-based LLM approaches reach saturation between 11 and 25 reports, with the Log4j campaign requiring 25 reports with LLMs. At saturation, both approaches stabilize with a precision range of 0.4 to 0.8.


For \textbf{Recall}, all approaches rise steeply within the first 5 to 10 reports, with most saturating after 10 to 15 reports. However, decoder-based LLMs on Log4j and XZ Utils campaigns continue to improve until 22 to 30 reports. At saturation, LLMs achieve the highest recall across campaigns—0.87 for Log4j, 0.85 for XZ Utils, and 0.75 for SolarWinds. NER achieves moderate recall, ranging from 0.75 for Log4j to 0.35 for XZ Utils, while encoder-based classification achieves the lowest recall, ranging from 0.5 for Log4j to 0.33 for XZ Utils.


For \textbf{F1-score}, saturation typically occurs after 5 to 13 CTI reports, depending on the extraction approach and attack campaign. No single approach consistently dominates across all campaigns. For example, in the Log4j campaign, NER achieves the best F1 score and reaches saturation after 5 to 10 reports, outperforming both encoder-based classification and decoder-based LLM approaches. In contrast, for the XZ Utils and SolarWinds campaigns, the decoder-based LLM performs best, saturating after 6 to 13 reports. Despite these improvements, the maximum F1-score for the XZ Utils campaign ($\approx$0.53) remains lower than even the lowest F1 score observed for the Log4j campaign ($\approx$0.61).

\begin{boxK}
\vspace{-3mm}
\textit{\autoref{RQSaturation}:} While all approaches generally saturate within 5–15 CTI reports, the optimal approach at saturation depends on the performance metric and attack campaign. NER achieves perfect precision after a single report, whereas decoder-based LLM achieves the highest recall but requires up to 30 reports. 
\vspace{-6mm}
\end{boxK}

\begin{table}[htbp]
\centering
\small
\caption{Cliff's $\Delta$ comparing pre- vs.\ post-saturation 
report characteristics. *** $p$<0.001, ** $p$<0.01, 
* $p$<0.05, ns = not significant.}
\label{tab:characteristics}
\setlength{\tabcolsep}{3pt}
\begin{tabular}{@{}llrrrr@{}}
\toprule
\textbf{Campaign} & \textbf{App.} & \textbf{Words} & \textbf{Sents.} & \textbf{Readab.} & \textbf{Days} \\
\midrule
\multirow{3}{*}{SolarWinds}
 & NER   & 0.84*** & 0.76**  & 0.16ns  & -0.30ns \\
 & Class.& 0.89*** & 0.89*** & 0.28ns  & -0.28ns \\
 & LLM   & 0.94*** & 0.88*** & 0.12ns  & -0.21ns \\
\midrule
\multirow{3}{*}{Log4j}
 & NER   & 0.68**  & 0.62*   & -0.54ns & 0.47*   \\
 & Class.& 0.95*** & 1.00*** & -0.23ns & 0.62**  \\
 & LLM   & 0.73*** & 0.64**  & -0.32ns & 0.20ns  \\
\midrule
\multirow{3}{*}{XZ Utils}
 & NER   & 0.31ns  & 0.24ns  & -0.05ns & -0.28ns \\
 & Class.& 0.76*** & 0.61**  & 0.39*   & 0.46*   \\
 & LLM   & 0.48*   & 0.53*   & 0.26ns  & 0.30ns  \\
\bottomrule
\end{tabular}
\end{table}

\subsection*{\reuseRQ{RQCTICharacteristics} \RQCTICharacteristics}

To answer \autoref{RQCTICharacteristics}, for each attack campaign and the best-performing method from each approach, we compare pre-segment and post-segment reports across five features using the Mann-Whitney U test, testing whether pre-segment reports score higher than post-segment reports, with effect sizes reported as Cliff's Delta  (described in \S~\ref{sec:performance_saturation}). ~\autoref{tab:characteristics} summarizes the results and the exact numbers for each feature of CTI reports in the supplementary materials.



\textbf{Report Length (Word Count and Sentence Count).} Pre-segment reports have a significantly higher number of word and sentence counts than post-segment reports across all three approaches for SolarWinds and Log4j. XZ Utils follows the same trend except for NER ($\Delta$ = 0.31, $p$ = 0.131), where the two groups show comparable lengths.
\textbf{Readability.} Readability shows no consistent directional difference between pre-segment and post-segment reports. Cliff's $\Delta$ ranges from $-$0.54 to 0.39 across campaigns and approaches, with the mix of positive and negative values indicating the absence of a systematic trend. The only statistically significant result is XZ Utils under encoder-based classification ($\Delta$ = 0.39, $p$ = 0.040).

\textbf{Publication Date.} Publication date shows no consistent directional pattern across campaigns. For Log4j, pre-segment reports under NER and encoder-based classification have significantly more days since disclosure ($\Delta$ = 0.47 and 0.62, $p$ < 0.05), but this pattern does not hold for decoder-based LLM. For SolarWinds, delta values are negative across all three approaches, indicating no advantage for earlier or later reports.
\textbf{Vendor.} Pre-segment reports are consistently dominated by government cybersecurity agencies and established commercial CTI vendors. For the SolarWinds campaign, CISA and Google account for $\approx$60\% of pre-segment reports across all three approaches, while post-segment reports are largely composed of SolarWinds' own disclosures and secondary sources. For Log4j, CISA and Apache collectively represent $\approx$65\% pre-segment reports, whereas post-segment reports span over a dozen vendors, including Openwall mailing list posts and Fedora. We see the same pattern for XZ Utils. 
\begin{boxK}
\vspace{-2mm}
\textit{\autoref{RQCTICharacteristics}:} CTI reports that are longer, contain technical content, and are from certain vendors consistently enable higher ATT\&CK technique extraction performance. In contrast, readability and publication date show no consistent correlation with technique extraction performance.
\vspace{-2mm}
\end{boxK}

\section{Discussion}
\label{sec:discussion}
\textbf{Security researchers should aggregate CTI reports from multiple sources rather than relying on a single report for ATT\&CK technique extraction.} Our results from ~\autoref{RQEffectiveness} show that no single CTI report captures the full ATT\&CK techniques of an attack campaign, and aggregating predictions across multiple reports consistently improves technique coverage on NER and decoder-based LLM approaches. From~\autoref{RQSaturation}, we also observe that saturation is typically reached within 5 to 15 reports with a threshold value of 0.005 in most cases, suggesting that analysts do not need to collect every available report. However, for the XZ Utils campaign, we found that the LLM saturates after 22 reports, with a threshold of 0.001. From \autoref{RQCTICharacteristics}, we found that overall, more than 60\% of the pre-segment reports are come from the government agencies (e.g., CISA) or established commercial CTI vendors (e.g., Google, Apache). 

\textbf{Security practitioners should select extraction approaches based on their analysis goals.} Results from~\autoref{RQEffectiveness} show that NER achieves perfect precision but the lowest recall, while decoder-based LLMs achieve the highest F1-score and recall. Such tradeoffs have direct implications for different security roles. For example, Security Operations Center (SOC) analysts and incident responders, who prioritize reducing false alerts and investigation overhead, may prefer NER. In contrast, threat intelligence analysts and detection engineers, who prioritize comprehensive technique coverage to understand attacker behavior, may prefer LLM-based extraction despite higher false-positive rates.


\textbf{Security analysts should prioritize CTI reports with rich technical content over readability.} Our~\autoref{RQCTICharacteristics} shows that report length and technical details have a significant effect on extraction performance, while the readability score has no significant effect. Readability scores penalize the inclusion of hashes, IP addresses, and domain-specific technical terms due to the long, complex, and unknown tokens they contain. So, a lower readability score may indicate a higher concentration of technical details rather than poor writing. Analysts curating CTI corpora should therefore favor longer, technically detailed reports, even if they appear harder to read.

\textbf{Researchers should complement ATT\&CK technique extraction with mitigation-level analysis to better characterize attack campaigns.}~\autoref{RQMitigationGap} shows that the best-performing automated extraction method achieves $\approx$ 90\% ATT\&CK technique coverage. However, mapping these predicted techniques to P-SSCRM controls reveals that only 77\% of the required controls are covered. 
This demonstrates that even a relatively small number of missed ATT\&CK techniques ($\approx$10\%) can generate substantial gaps in control coverage ($\approx$23\%), highlighting the importance of analyzing both ATT\&CK techniques and associated controls to fully characterize attack campaigns.

\vspace{-2mm}
\section{Threats to Validity}
\label{sec:threats_to_validity}
\textbf{Construct validity.} Mapping ATT\&CK techniques to P-SSCRM controls introduces potential subjectivity. We mitigate this by extending the four independent mapping strategies proposed by Hamer et al.~\cite{hamer2026closing} with three additional independent manual mappings to provide diverse perspectives and convergence.
\textbf{Internal validity.} Training on AnnoCTR and testing on CTIfecta introduces potential distributional differences and missing technique coverage. We mitigate this by augmenting the training data with MITRE ATT\&CK-derived samples for techniques absent in AnnoCTR, preserving the long-tail distribution, and by leveraging domain-specific transfer learning~\cite{lin2022attack, aghaei2022securebert} and data augmentation~\cite{gao2023benefits, ruiz2025synthcti} to improve cross-dataset generalization.
\textbf{External validity.} Our findings may not generalize to other types of attack campaigns beyond SolarWinds, Log4j, and XZ Utils. Additionally, the dataset is limited to CTI reports collected for these three campaigns, which is mitigated by the dataset authors~\cite{hamer2026closing} through three sampling strategies designed to achieve theoretical saturation~\cite{baltes2022sampling}.
\vspace{-3mm}
\section{Conclusion}
\label{sec:conclusion}
We evaluated 29 state-of-the-art ATT\&CK technique extraction methods spanning three approaches using multiple CTI reports from the SolarWinds, Log4j, and XZ Utils campaigns. We found that aggregating multiple reports gives better performance than a single report, with most approaches reaching performance saturation within 10 to 15 reports. Reports that are longer and include more technical details contributed the most to the saturation. Despite the improvement, performance remains limited; $\approx$33.3\% of misclassifications occur between techniques sharing the same MITRE tactic. Furthermore, extraction errors disproportionately propagate into controls: the best-performing method misses only 10\% of ATT\&CK techniques but results in a 23\% gap in controls. Future work should explore campaign-level multiple reports with technical details and evaluate multi-modal extraction pipelines that process command traces alongside plain text, evaluating them against both ATT\&CK techniques and control identification.



\section{Data availability}
We release the dataset and replication package at \url{https://figshare.com/s/9ad0a0a0aa4d390b7241}.

\bibliographystyle{ACM-Reference-Format}
\bibliography{ref}

\end{document}